\documentclass[10pt, conference, compsocconf]{IEEEtran}
\ifCLASSINFOpdf
 \usepackage[pdftex]{graphicx}
\else
\fi
\usepackage{array}
\usepackage{amsmath}
\usepackage{amsfonts}
\usepackage{amssymb}

\usepackage[mathscr]{euscript}
\newcommand{\X}{\mathcal{X}}
\newcommand{\DXP}{$d_{\X}$-privacy \,}
\newcommand{\DXPns}{$d_{\X}$-privacy}
\newcommand{\DX}{$d_{\X}$ \,}
\newcommand{\DXns}{$d_{\X}$}
\newcommand{\eDXPrivate}{$\varepsilon\cdot d_{\X}$-private \,}
\newcommand{\DXPrivate}{$d_{\X}$-private \,}
\newcommand{\mech}{\mathscr{M}}

\usepackage{color}

\usepackage[hyphenbreaks]{breakurl}
\usepackage[hyphens]{url}

\begin{document}
%
\title{Metric-based local differential privacy for statistical applications}


\author{\IEEEauthorblockN{M\'ario S. Alvim\IEEEauthorrefmark{1},
Konstantinos Chatzikokolakis\IEEEauthorrefmark{2}\IEEEauthorrefmark{3},
Catuscia Palamidessi\IEEEauthorrefmark{4}\IEEEauthorrefmark{3} and Anna Pazii\IEEEauthorrefmark{3}\IEEEauthorrefmark{4}}
\IEEEauthorblockA{\IEEEauthorrefmark{1}UFMG, Belo Horizonte, Brazil}
\IEEEauthorblockA{\IEEEauthorrefmark{2}CNRS, France}
\IEEEauthorblockA{\IEEEauthorrefmark{3}Ecole Polytechnique, University of Paris Saclay, France}
\IEEEauthorblockA{\IEEEauthorrefmark{4}INRIA, University of Paris Saclay, France}}



\maketitle

\begin{abstract}
Local differential privacy (LPD) is a distributed variant of differential privacy (DP) in which 
the obfuscation of the sensitive information is done at the level of the individual records, 
and in general it  is used to sanitize data that are collected for statistical purposes.  
LPD has the advantage it does not need to assume a trusted third party. 
On the other hand LDP in general requires more noise than DP to achieve the same level of protection, 
with negative consequences on the utility. In practice, utility becomes acceptable only on very large collections of data, and 
this is the reason why LDP is especially successful among big companies such as Apple and Google, which can count on a 
huge number of users.  
In this paper, we propose a variant of LDP suitable for metric spaces, such as location data or energy consumption data, 
and we show that it provides a much better utility for the same level of privacy. 

\end{abstract}


%
\IEEEpeerreviewmaketitle

\section{Introduction}
With the ever-increasing use of internet-connected devices, such as computers, IoT appliances (smart meters, home monitoring devices), 
and GPS-enabled equipments (mobile phones, in-car navigation systems),  
personal data are collected in larger and larger amounts, and then stored and manipulated  
for the most diverse purposes. For example, the web browsing history can be used for profiling  the user and sending him targeted publicity. 
Power-consumption data from smart meters can be analyzed to extract typical daily
consumption patterns in households
 \cite{Hino:13:TOSG}, 
or to identify the right customers to target for demand response programs
\cite{Chelmis:15:BD}.
Location data can be used to find the most frequented public areas (for instance, to deploy hotspots)
\cite{Zheng:2009:WWW}, 
or to provide traffic information~\cite{Hull:2006:ENSS}.  

Undeniably, the Big Data technology   provides enormous benefits to
individuals and society, ranging from helping the scientific progress to improving the quality of service. 
On the other hand, however, the collection and manipulation 
of personal data raises alarming privacy issues. 
Not only the experts, but also the population at large are becoming increasingly aware of the risks, 
due to the repeated cases of violations and leaks that keep appearing on the news. 
It is particularly disturbing when personal data are collected without the user's consent, or even awareness. 
For instance, in 2011 
it was discovered that the iPhone was storing and collecting
location data about the user, syncing them with iTunes and transmitting them
to Apple, all without the user's knowledge~\cite{Bilton:11:NYT}.
More recently, the Guardian has revealed, on the basis of the documents provided by Edward Snowden, 
that the NSA and the  GCHQ have been using certain smartphone apps, such as the wildly popular
Angry Birds game, to collect users' private information  such as age, gender and
location~\cite{Ball:14:Guardian}, again without the users' knowledge. 

Another related problem is that users  often do not have the possibility to control the precision and the amount 
of personal information that is being exposed. 
For instance the Tinder application was found sharing the exact
latitude and longitude co-ordinates of users as well as their birth dates and
Facebook IDs \cite{Seward:2013:QUARTZ}, and even after the initial problem was fixed, 
it was still  sharing more accurate location data than intended, as users could be located to within 100
feet of their present location \cite{Dredge:2014:TG}. 
The leakage of {\it precise} personal information is particularly problematic, especially 
when considering that  various kinds of personal data from different
sources can be linked and aggregated into a user profile
\cite{Hasan:13:BD,Schiaffino:09:AI}, 
 and can fall in the hands of malicious parties.

Until recently,  the most popular and used data sanitization technique was  
anonymization (removal of names) or  slightly more sophisticated variants like $k$-anonymity \cite{Samarati:01:TKDE} ensuring indistinguishability within groups of at least $k$ people, 
and $\ell$-diversity, ensuring a variety of values for the sensitive data within the same group \cite{Machanavajjhala:07:TKDD}. 
Unfortunately, these techniques have been proved unable to provide an acceptable level of protection, as several works have shown that individuals in 
anonymized datasets can be re-identified with high accuracy, and their personal information exposed (see for istance \cite{Narayanan:08:SSP,Narayanan:09:SSP}).

In the meanwhile, \emph{differential privacy} (DP), has emerged and imposed itself as a convincing alternative to anonymity. Together with its distributed version \emph{local differential privacy}, it represents the cutting-edge of research on privacy protection. 

DP was developed in the area of statistical databases, and it aims at protecting the individuals' data while allowing to 
make public the aggregate information~\cite{Dwork:06:TCC}. This is obtained by adding controlled noise to the query outcome, in such a way that  the data of a single individual will have a negligible impact on the reported answer. 
More precisely, let $\mech$ be a (noisy) mechanism 
for answering a certain query on a generic database $D$, and let $P[{\mech}(D)\in S]$ denote the probability that the answer given by $\mech$ to the query, on $D$, is in the (measurable) set of values $S$. 
We say that $\mech$
is $\varepsilon$-differentially private  if   for every pairs of \emph{adjacent}  databases $D$ and $D'$ (i.e., differing only for the value of a single individual), and for every measurable set  $S$, we have $ P[{\mech}(D)\in S]\leq e^{\varepsilon} \cdot P[{\mech}(D')\in S]$.
DP  has two important advantages with respect to other approaches: (a) it is independent from the side-information of the adversary, thus a differentially-private mechanism can be designed without  taking into account the context in which it will have to operate, and (b)  it is compositional, i.e., if we combine the information that we obtain by querying two differentially-private mechanisms, the resulting mechanism is also differentially-private.   
Furthermore, (c) differentially-private mechanisms usually provide a good trade-off between utility and privacy, i.e., they preserve the privacy of the individuals 
without destroying the utility of the collective data.                                                                                                                                                                                                                                                                                                                                                                                                                                                                                                                                                                                                   

Local differential privacy (LDP) is a distributed variant of differential
privacy in which users obfuscate their personal data by themselves, before sending them 
to the data collector~\cite{Duchi:13:FOCS}. 
Technically, an obfuscation mechanism
$\mech$ is locally differentially private with privacy level $\varepsilon$ if for every pair of input values
$x, x'\in \X$ (the set of possible values for the data of a generic user), and for every measurable set  $S$,  we have  
$ P[{\mech}(x)\in S]\leq e^{\varepsilon} \cdot P[{\mech}(x')\in S]$.
The idea is that the user provides $\mech(x)$  to the  data collector, and not $x$.
In this way, the data collector  can only gather, stock and analyze  the obfuscated data. 
Based on these
he can infer statistics (e.g., histograms, or
heavy hitters \cite{Qin:16:CCS}) of the original data. 
LPD implies DP on the collected data,  and has the same advantages of independence from the side-information and compositionality. 
Furthermore, with respect to the centralized model, it has the further advantages that (a) each user can choose the level of privacy he wishes, 
(b) it does not need to assume a trusted third party, and (c) since all stored  records are individually-sanitized ,   
there is no risk of privacy breaches due to malicious attacks. 
LDP is having a considerable impact, specially after large companies such as Apple and Google 
have started to adopt it for collecting the data of their customers for statistical purposes 
\cite{Erlingsson:14:CCS}. 

The disadvantage of LDP is that it can spoil substantially the utility of the data. 
Even in those cases where the trade-off with utility is most favorable, namely 
the statistical applications, it is usually necessary to have a huge collection of data in order 
for the statistics to be significant.
Fortunately, however, the data domains are often equipped with structures  that could be exploited to improve utility. 
In these notes, we focus on data domains that are provided with a notion of distance. 
This is the case, for instance, of location data, energy consumption in smart meters, age and weight in medical records, etc. 
Usually, when these data are collected for statistical purposes, the accuracy of the distribution 
is measured also with respect to the  same notion of distance. 
In such scenarios, \emph{we argue that the trade-off between privacy and utility 
can be greatly improved by exploiting the 
concept of approximation intrinsic in metrics}.

Following  this intuition, we propose a variant of local differential privacy based 
on the notion of \DXP intoduced in \cite{Chatzikokolakis:13:PETS}. 
An obfuscation mechanism ${\mech}$ is 
\eDXPrivate  if for every pair of input values
$x, x'\in \X$, and for every measurable set  $S$,  we have  
$ P[{\mech}(x)\in S]\leq e^{\varepsilon\cdot d_X(x,x')} \cdot P[{\mech}(x')\in S]$.
In other words, \DXP relaxes the privacy requirement
by allowing two data to become more and more distinguishable as their distance increases. 
Thus, it allows the adversary to infer some approximate information about the true
value, but it does not allow him to infer the exact true value.
As explained in  \cite{Chatzikokolakis:13:PETS}, \DXP can be implemented by using 
an extended notion of Laplacian noise, or of geometric noise. 

The original motivation  for the notion of  \DXP  was for real-time punctual applications. In particular, the instance of  \DXP in which  \DX is the geographical distance  has been used in the context of location privacy, under the name of \emph{geo-indistinguishability}, 
to protect the location of the user during the interaction with location-based services (LBSs) \cite{Andres:13:CCS,Chatzikokolakis:17:POPETS}. 
These are services that provide 
the user with certain desired information which depends on the location communicated by the user, like for instance points of interest (POI) near the location. 
The idea is that the user does not need to communicate his exact coordinates, 
an approximate location should suffice to obtain the requested information without too much degradation of the quality of service. 

Geo-indistinguishability has been quite successful, and its implementation via the Laplacian  mechanism has been adopted as the basis or as a component of several tools and frameworks for location privacy, including:  
LP-Guardian \cite{Fawaz:14:CCS}, LP-Doctor \cite{Fawaz:15:USENIX},   STAC \cite{Pournajaf:15:ICAGIS},  Location Guard \cite{location-guard-github},   and the SpatialVision QGIS plugin \cite{QGIS}. 
Here, we want to show that geo-indistinguishability, and more in general \DXPns, can also be used to protect privacy when collecting data for  statistical purposes, 
and that if the statistics are distance-sensitive, then \DXP preserves the utility of the data better than the standard LDP methods.

In the rest of these notes, we will discuss the improvement on trade-off utility-privacy compared to standard LDP methods, and 
show some experimental results based on the Gowalla dataset~\cite{Cho:11:SIGKDD,Gowalla}. 
For simplicity we will restrict the analysis to the case of discrete metric spaces.
We will consider, in particular, the mechanisms of K-ary Randomized Response (K-RR)~\cite{Kairouz:16:JMLR} for LPD, and 
the (discretized) Laplacian and geometric mechanisms for \DXPns.

\section{Utility}
We consider a  notion of utility suitable  for statistical applications. 
The scenario is the following: let $\mathcal{X}$ (the universe) be a set of secrets, endowed with a notion of distance \DXns, and let $\mathbb{D}\mathcal{X}$ be the set of distributions on $\mathcal{X}$. 
Let $D$ be an unsanitized dataset on $\mathcal{X}$, namely a multiset of elements of $\mathcal{X}$ (i.e.,  an histogram), determining a distribution $\pi\in \mathbb{D}\mathcal{X}$. 
Assume that each individual element $x$ in $D$ gets sanitized 
by injection of noise, thus producing a noisy dataset $\hat{D}$. 
From $\hat{D}$ we then try to reconstruct the distribution $\pi$ as well as we can, assuming that we only know  $\hat{D}$
and the mechanism $\mech$ for noise injection. 

In order to reconstruct as precisely as possible the original $\pi$, 
we propose to use  the Expectation-Maximization (EM) method \cite{Agrawal:05:ICMD},  also known as Iterative Bayesian Update, 
which iteratively estimates the distribution until convergence to a fixed point. 
The feature of this
method is that the final estimate (converged value)  is
equal to the Maximum Likelihood estimate in the
probability simplex, and it is shown in \cite{Agrawal:05:ICMD} that  it significantly outperforms the other known techniques
like the matrix inversion method.

Let  $\hat{\pi}\in\mathbb{D}\mathcal{X}$ be the output of the EM method. Intuitively, the \emph{utility loss} with respect to the original database should reflect the \emph{expected difference   
between the statistical properties based on the noisy data and those based on the real data}. 
To formalize the notion of \emph{expectation}, we can regard $\mech$, in abstract terms,  as a device that inputs $\pi$ and output a set of possible 
distributions $\hat{\pi}_1, \hat{\pi}_2, \ldots \hat{\pi}_i\ldots $, each with a certain probability  $p_1, p_2, \ldots p_i\ldots$. 
In other words, $\mech$ can be seen as a transformation that associates to each $\pi$ a  function $\Delta$ 
which assigns a probability mass to every distribution,  i.e,  $\Delta(\hat{\pi}_i)= p_i$. 
This type of  functions $\Delta$ are called  \emph{hyperdistributions} in 
\cite{McIver:10:ICALP,Alvim:14:CSF}. Note that also $\pi$ can be seen as a hyperdistribution: it is the function that assigns $1$ to $\pi$, and 
we will denote it by $[\pi]$.

Concerning the \emph{difference between statistical properties}: in very general terms, we can represent a statistical property as a functions  $f:\mathbb{D}\mathcal{X}\rightarrow \mathbb{R}$, where $\mathbb{R}$ is the set of reals. We want to capture as many $f$'s as possible, but it is reasonable to assume that the difference between $f(\pi)$ and $f(\pi')$ must be bound by the distance between the distributions $\pi$ and $\pi'$, for some ``reasonable'' notion of distance $d_{\mathbb{D}\mathcal{X}}$. In other words, we want to avoid that negligible differences on the distributions may produce unbound differences in the statistics. For this reason, we restrict the statistics of interest, $\mathscr{F}\subseteq (\mathbb{D}\mathcal{X}\rightarrow \mathbb{R})$, to be the set of $1$-Lipshitz\footnote{The requirement of $1$-Lipshitz is not really essential, it could be $k$-Lipshitz for an arbitrary $k$. The important constraint is that the difference on the statistics is bound in some uniform  way by the difference on the distributions.}
functions with respect to $d_{\mathbb{D}\mathcal{X}}$. Hence,  $\mathscr{F}=\{f:\mathbb{D}\mathcal{X}\rightarrow \mathbb{R}\mid f \mbox{ is $1$-Lipshitz w.r.t. } d_{\mathbb{D}\mathcal{X}}\}$. Finally, since we want to abstract from the peculiarity of any particular statistics, we will  consider the \emph{maximum difference} induced by the noise on all the statistics in $\mathscr{F}$. Summarizing, we can define the utility loss as: 
\begin{equation}\label{eqn:utility_loss}
\mathscr{UL}(\mathscr{M},\pi,d_{\mathbb{D}\mathcal{X}})  \;=\; \max_{f\in\mathscr{F}} \mid \sum_{\hat{\pi}}\Delta(\hat{\pi}) f(\hat{\pi})- f(\pi)\mid
\end{equation}
where $\Delta=\mathscr{M}(\pi)$. 
It is worth noting that the rhs of this definition is the distance between $\Delta$ and $[\pi]$ obtained as the \emph{Kantorovich lifting} of $d_{\mathbb{D}\mathcal{X}}$, which we will denote as $K(d_{\mathbb{D}\mathcal{X}})(\Delta,[\pi])$.  
Following the same intuition, we can define the distance $d_{\mathbb{D}\mathcal{X}}$  as the \emph{Kantorovich lifting} of \DXns, thus establishing a link with the ground distance on the domain of secrets. 

\section{Tuning privacy}
The notions of LDP and of \DXP both depend on  privacy parameters $\varepsilon$'s, but these $\varepsilon$'s do not represent the same level of privacy  in the two definitions. They are   not even of the same type: the $\varepsilon$ in LDP is a pure number, while the $\varepsilon$ in \DXP is 
the converse of a distance. 
Therefore, in order to compare the utility of an LDP mechanism $\mech$ with that of a \DXPrivate  mechanism $\mech'$, we have first to tune their privacy parameters so to ensure that $\mech$ and $\mech'$ provide the same privacy protection. 
To this end, we consider the notion of location privacy proposed in \cite{Shokri:17:TPS}, defined as expected distance between the reported location and 
the real location.~\footnote{The definition in \cite{Shokri:17:TPS}  also takes into account the knowledge of the prior, and the possibility to remap the reported location in the most most likely one according to the additional information provided by the prior. In our case, we want a definition that does not depend on the knowledge of the prior (since $\pi$ is supposed to be unknown), and without  the prior  information, for the mechanisms we consider, the most likely location is always the reported one. Hence we do not need remapping.}
Generalizing  to \DXPns, we require that $\mech$ and $\mech'$ give the same expected distance between $x\in D$ and the corresponding reported datum. Namely:
\begin{equation}\label{eqn:tuning_privacy}
\begin{array}{c}
\sum_{x,y\in \mathcal{X}}\pi(x) P[ \mathscr{M}(x)=y] \mbox{\DXns}(x,y) \\ =  \\  Ed \\=\\ \sum_{x,y\in \mathcal{X}}\pi(x) P[ \mathscr{M}'(x)=y] \mbox{\DXns}(x,y)  
\end{array}
\end{equation}
where $Ed$ represents the desired level of protection, expressed in terms of the expected distance of the reported location from the real one. 

\section{The mechanisms}
We now recall the definition of the K-ary RR mechanism~\cite{Kairouz:16:JMLR}, representative of LPD, and 
the Laplacian and geometric mechanisms, representative of  \DXPns.
\subsection{The  Laplacian mechanism}
The Laplacian mechanism $\mathscr{M}_L$ is used when $(\mathcal{X},\mbox{\DXns})$ is a continuous metric space. Given a  real location x, it reports a location $y$ with a probability density function (pdf) defined as:
\[ dP_{x}(y) = \lambda_L \, e^{-\varepsilon\cdot \mbox{\DXns}(x,y)}\]
where $\lambda_L$ is a normalization factor. 

In case we want to work in a discrete setting, we can first discretize the metric space by  partitioning   $\mathcal{X}$ into cells 
and defining the distance between two cells as the \DX between the centers of the cells. Then we can discretize the mechanism by defining the probability of a cell $C$ as the probability mass 
obtained by the integration of the pdf over  $C$. 

\subsection{The geometric mechanism}
 
	The   geometric  mechanism $\mathscr{M}_G$ is used  when $(\mathcal{X},\mbox{\DXns})$ is a discrete metric space.  It is defined similarly to  Laplacian  mechanism, with the exception that now $x$ and $y$ represent discrete locations, and we have a (discrete) 
probability distribution rather than a pdf: 
 \[ P[\mathscr{M}_G(x) = y] = \lambda_G \, e^{-\varepsilon\cdot \mbox{\DXns}(x,y)}\]
where $\lambda_G$ is a normalization factor. 

If  $(\mathcal{X},\mbox{\DXns})$ is the result of a discretization of a continuous metric space, then the discretized laplacian is very similar, but not identical to the geometric mechanism.  

\section{The  K-RR mechanism}
The K-RR mechanism, aka \emph{flat} mechanism, $\mathscr{M}_F$  is one of the simplest LPD mechanisms. The idea is that the result of the sanitization is a bit more likely to be the true value $x$ than  any other value in the domain (taken individually), and that on the other values the probability is distributed uniformly :
\begin{center}
$P[\mathscr{M}_F(x) = y] =  \left\{ 
\begin{array}{ll}
  \frac{e^{\epsilon}}{|X|-1+e^\epsilon}  & \mbox{if }  y = x \\[2ex]
  \frac{1}{|X|-1+e^\epsilon}  & \mbox{if } y \neq x
\end{array}
\right.
$
\end{center}

 In \cite{Kairouz:16:JMLR}, the k-RR mechanism has been shown to be optimal in the low privacy regime for a large class of information theoretic utility functions.

\section{Experimental results}
In this section we compare the utility of 
the privacy mechanisms $\mathscr{M}_F$, $\mathscr{M}_G$ and the (discretized) $\mathscr{M}_L$ introduced in previous section, using a
distribution derived from the Gowalla dataset, which contains  location data (check-ins) relative to a certain population of users. 

\begin{figure}
\begin{center}
\includegraphics[width=0.4\textwidth]{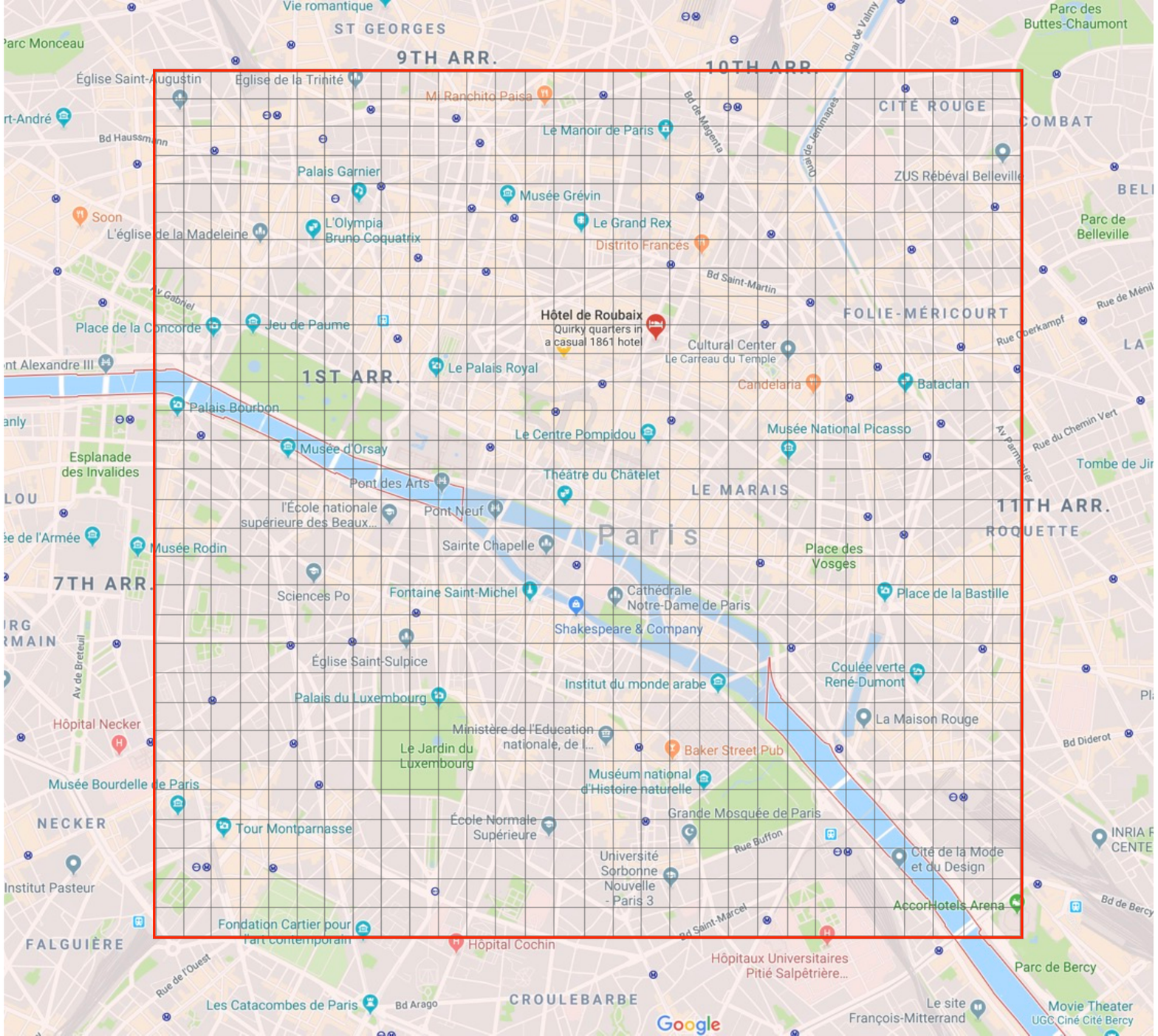} 
\end{center}
\caption{
The area of Paris consider for evaluating the utility of the three mechanisms.  }
\label{fig:area}
\end{figure}

We  consider an area of $4.5$ Km $\times$ $4.5$ Km in Paris, centered in $5$ Boulevard de S\'ebastopol, near Le Halles. 
We discretize that area by considering a grid of $30\times 30$ cells, so that every cell is $150$ m $\times 150$ m, see Figure~\ref{fig:area}.
These cells represent the elements of 
$\mathcal{X}$, and the distance \DX is defined as the geographic distance between the centers of the cells.

We consider $750$ check-ins from Gowalla in this area, selected randomly, and we consider the multiset $D$ obtained by counting the number of check-ins in each cell. Let $\pi$ be the corresponding distribution. 

We now tune the privacy parameters of $\mathscr{M}_F$, $\mathscr{M}_G$ and  $\mathscr{M}_L$ so that 
the expected distance $Ed$ of the reported location from the real one is the same for all of them (cfr. Requirement \eqref{eqn:tuning_privacy}). 
We set $Ed$  to be $3$ times the size of a cell, namely $450$ m. The values of $\varepsilon$ that we derive are: $8.24797$ for $\mathscr{M}_F$, $0.00398441$ for $\mathscr{M}_G$, and 
 $0.00404249$ for $\mathscr{M}_L$.

\begin{figure}
\begin{center}
\includegraphics[width=0.4\textwidth]{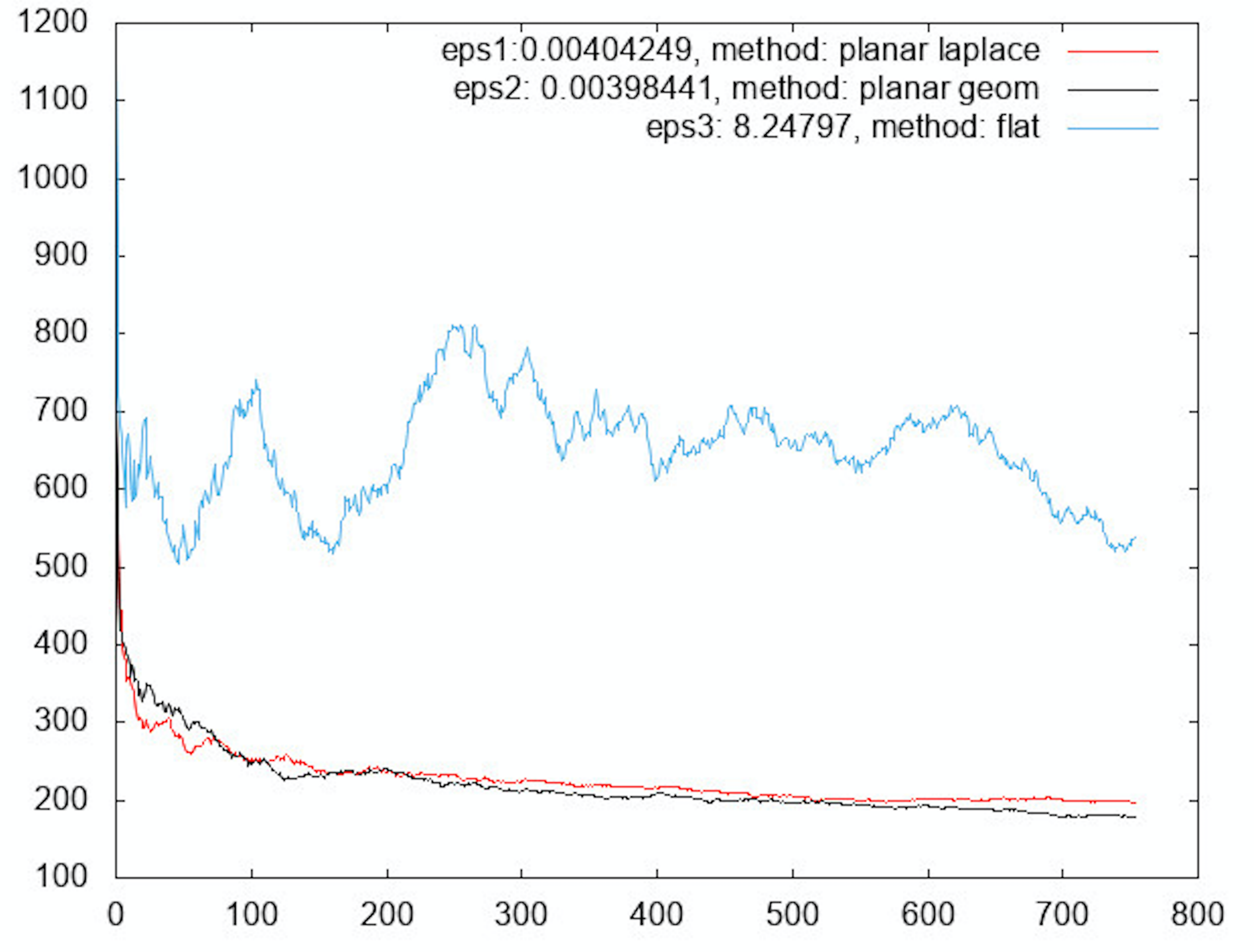} 
\end{center}
\caption{
The utility loss for an increasing sequence of datasets taken from the Gowalla check-ins in the area illustrated in Figure~\ref{fig:area}. The
numbers in the horizontal axis represent the number of check-ins $\times$ $10$. The
numbers in the vertical axis represent the distance, expressed in meters.}
\label{fig:evaluation}
\end{figure}

In order to compute the utility loss for these three mechanisms, 
we use the well-known fact that the Kantorovich distance is equal to the earth mover's distance (EMD) also known as the Wasserstein metric.
Namely, we can equivalently rewrite  \eqref{eqn:utility_loss} as
\begin{equation}\label{eqn:utility_loss_dual}
\mathscr{UL}(\mathscr{M},\pi,d_{\mathbb{D}\mathcal{X}})  \;=\; \min_\alpha  \sum_{\hat{\pi}\in \mathit{dom}(\Delta)}\alpha(\hat{\pi},\pi) d_{\mathbb{D}\mathcal{X}}(\hat{\pi},\pi)
\end{equation}
where $\Delta= \mathscr{M}$,  $\alpha$ ranges on the couplings that have as marginals $\Delta$ and $[\pi]$, 
and  $d_{\mathbb{D}\mathcal{X}}=K(\mbox{\DXns})$.
In conclusion, to determine the utility loss we need to compute the expectation of the  Kantorovich distance between the reported location and the real location. 
We have done it for an increasing sequence of  dataset $\emptyset \subseteq D_1\subseteq D_2 \subseteq \ldots \subseteq  D_n \subseteq  \ldots  \subseteq  D$ constructed incrementally by adding each time $10$ elements from the $750$ check-ins, selected randomly, until all of them are inserted. The results are reported in Figure~\ref{fig:evaluation}.

As we can see from this figure, the geometric and the discretized Laplacian have similar utility, and perform much better (in terms of utility) than the flat mechanism. 
In fact, if we consider statistics that are somehow coherent with the ground distance \DXns, then the fact that the geometric and the discretized Laplacian tend to assign a negligible probability to locations that are far away from the real one means that those locations do not significantly contribute to the loss of utility. 
In contrast, the flat mechanism treats in the same way all locations, independently from their distance from the real one. Consequently there are several locations that are far away and still carry a significant probability mass, thus taking a heavy toll on the utility.

Furthermore, we can see that the utility loss of the Flat mechanisms, although showing a tendency to diminish as the numbers of check-ins increases, 
it does so very slowly. Finally, at the beginning (for less than $2000$ check-ins) the behavior of the flat mechanism is extremely unstable.  This is due again to the fact that the reported locations tend to be scattered in the whole area with high probability, which determine high fluctuations especially at the beginning when the data are few, as the addition of new data can cause a big change in the distributions.

\section{Conclusion}
We have advocated the use of \DXP 
to protect privacy when data are collected for statistical 
purposes on domains of secrets 
endowed with a notion of distance, arguing that in such context 
\DXP offers a better trade-off between privacy and utility than traditional LPD methods. 
We have confirmed this claim by performing  experimental evaluations of the utility of 
\DXns-private mechanisms and LPD ones on real location data from the Gowalla dataset. The results show that 
the gap in terms of utility (for the same level of privacy) is actually quite significant. 


\section*{Acknowledgment}
This work has been supported by the ANR project REPAS and  
by the project Epistemic Interactive Concurrency (EPIC) from the
862 STIC AmSud Program. Ma\'{a}rio S. Alvim was supported by CNPq, CAPES, and FAPEMIG.



\bibliographystyle{IEEEtran}
\bibliography{IEEEabrv,../../../../Bibliographies/biblio_cat}
%
%
%

\end{document}